\documentclass[prl,twocolumn,superscriptaddress,showpacs,floatfix,amsmath,amssymb]{revtex4-1}
\usepackage{xcolor}
\usepackage[utf8]{inputenc}
\usepackage{amsmath}
\usepackage{amssymb}
\usepackage{graphicx}
\usepackage{gensymb}
\usepackage{graphics}
\usepackage[T1]{fontenc}
\usepackage{soul}

\begin{document}
\title{Light-tunable charge density wave orders in MoTe$_2$ and WTe$_2$ single layers}
\author{Giovanni Marini}
\affiliation{Graphene Labs, Fondazione Istituto Italiano di Tecnologia, Via Morego, I-16163 Genova, Italy}
\email{giovanni.marini@iit.it}
\author{Matteo Calandra} 
\affiliation{Department of Physics, University of Trento, Via Sommarive 14, 38123 Povo, Italy}
\affiliation{Graphene Labs, Fondazione Istituto Italiano di Tecnologia, Via Morego, I-16163 Genova, Italy}
\email{m.calandrabuonaura@unitn.it}
\begin{abstract}
By using constrained density functional theory modeling, we demonstrate that ultrafast optical pumping  unveils hidden charge orders in group VI monolayer transition metal ditellurides. We show that irradiation of the insulating 2H phases stabilizes multiple transient charge density wave orders with light-tunable distortion, periodicity, electronic structure and bandgap. Moreover, optical pumping of the semimetallic 1T$^{\prime}$ phases generates a transient charge ordered metallic phase composed of 2D diamond clusters. For each transient phase we identify the critical fluence at which it is observed and the specific optical and Raman fingerprints  to directly compare with future ultrafast pump-probe experiments. Our work demonstrates that it is possible to stabilize charge density waves even in insulating 2D transition metal dichalcogenides by ultrafast irradiation.
\end{abstract}
\maketitle

Designing and manipulating broken symmetry states with laser light is an appealing perspective as it can lead to the discovery of hidden ordered states and enable control over a broad range of  material properties \cite{10.1088/1361-648X/abfe21,doi:10.1021/jacs.9b10533}. 
Few layers transition metal dichalcogenides (TMDs) are an ideal class of materials for ultrafast investigations, mainly for two reasons. First, metallic dichalcogenides display the occurrence of competing orders such as charge density wave (CDW) phases and superconductivity. Second, insulating TMDs normally do not display charge ordering but have typical gaps in the $1-2.5$ eV range\cite{Rasmussen2015}, ideal for optical pumping . Inducing CDW in insulating dichalcogenides is important as it could lead to new low dimensional phases with unexpected topological and correlation properties. 

Particularly relevant are MoTe$_2$ and WTe$_2$  as the barrier existing between the 2H and the 1T$'$ phase is lower compared to the other compounds of the family~\cite{Qian1344}. The most stable MoTe$_2$ polytype is the 2H, however it has been shown that a transition towards the 1T$^{\prime}$ can be selectively activated in ultrathin layers by means of electrostatic doping~\cite{Wang2017}, tensile strain~\cite{doi:10.1021/acs.nanolett.5b03481} and laser irradiation~\cite{Cho625}. Conversely, WTe$_2$ is stable in the 1T$'$ polytype and the metastable 2H phase has been synthesized only recently~\cite{doi:10.1021/acs.inorgchem.0c02049,D0CS00143K}. 

Most of the work carried out on irradiated MoTe$_2$ concerns the irreversible 2H-1T$^{\prime}$ phase transition.
However, irreversible phase transitions are only a small part of the broken symmetry charge ordered states available after laser irradiation, as reversible transition towards transient phases can occur; these can be detected either by ultrafast X-ray diffraction at X-ray free electron laser facilities\cite{PhysRevLett.117.135501} or by pump-probe experiments to measure optical or Raman spectra after the electronic excitation\cite{Ferrante2018,PhysRevLett.80.185}.  

Phase transitions under ultrafast irradiation have been observed for conventional semiconductors under intense ultrafast irradiation\cite{Siders1999,Rousse2001} (non-thermal melting), in phase-change materials\cite{PhysRevLett.117.135501} and have been proposed to occur in  ferroelectrics \cite{PhysRevLett.123.087601}. However, in most of these cases, the ordered phase is present in the ground state and suppressed by irradiation. Very few detections of hidden CDW orders (i.e. the ordered phase is a transient state induced by light) have been devised or demonstrated\cite{Kogar2020}. Furthermore, it is unclear if hidden orders are present in single layer dichalcogenides and at what fluence they can be observed. 

In this work we demonstrate the occurrence of multiple charge ordered transient states in MoTe$_2$ and WTe$_2$ single layers after ultrafast irradiation of the 2H and 1T$^{\prime}$ phases in the high-excitation density regime ($n_e \geq 10^{14}$ $~e^-/$cm$^2$, where $n_e$ represents the photocarrier concentration (PC)). Moreover we show the occurrence of a light-induced bandgap closing in the 2H phases. Finally, we identify specific spectroscopic fingerprints of each phase that will allow its detection in future ultrafast experiments. We investigate the effect of laser irradiation within density functional theory (DFT). In the case of insulators, we suitably constrain the occupations of the Kohn-Sham conduction eigenstates in order to mimic the thermalized photocarrier population. This approach has been described in Ref.~\cite{PhysRevB.65.054302} and we refer to it as constrained DFT (cDFT). We implemented the cDFT technique as well as the calculation of forces and stress tensor in the presence of an electron-hole plasma within the Quantum ESPRESSO distribution\cite{QE,QE2}. In the case of metals and semimetals (1T$^{\prime}$ phases), the radiative electron-hole recombination is fast\cite{Schmitt1649,doi:10.1021/nl301035x}, thus laser excitation is simulated employing Fermi-Dirac occupations at high temperature.  All the simulation details are reported in the Supplemental Material (SM), which includes Refs.\cite{doi:10.1021/acs.jctc.6b00114,PBE,MP,Naylor_2017,C5CP01649E,doi:10.1063/1.1564060,doi:10.1063/1.1760074,PL,C5NR00383K,Chernikov2015,doi:10.1142/7184,doi:10.1021/nl503799t,doi:10.1021/acsanm.8b02008,PhysRevB.94.085429,doi:10.1021/nl502557g,C5CP06706E,2009CoPhC.180.2622A,KOKALJ1999176,Momma:db5098,PhysRevLett.107.015501,erben2020optical,Laturia2018,PhysRevB.96.075448,PhysRevLett.82.3296,PhysRevB.94.094114,C7NR07890K,C6RA23687A,C6NR00492J,Cao_2017,C5NR06098B,PhysRevB.99.235401}.

\begin{figure*}[t!]
\centering
\includegraphics[width=1\linewidth]{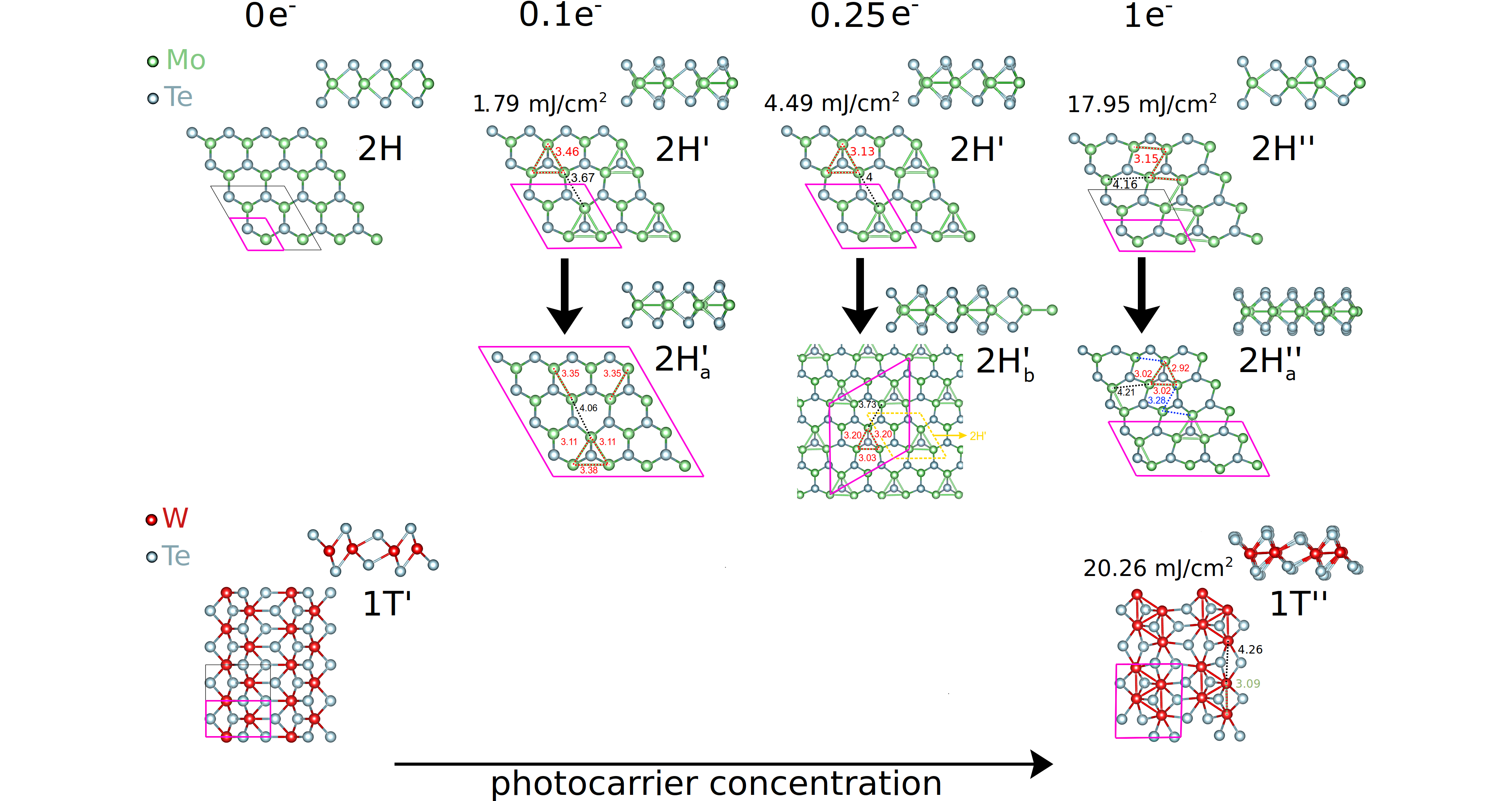}
\caption{Schematic representation of hidden orders in monolayer dichalcogenides. Upper panel: 2H-MoTe$_2$ as a function of the PC ($f.u.$). Mo clusterization is represented by green bonds. At $n_e=0.25~e^-/f.u.$ $(n_e=1~e^-/f.u.)$ the energy gain associated to the 2H $\rightarrow$ 2H$'$ (2H $\rightarrow$ 2H${''}$) distortion is $107.2$ (583.1) meV$/f.u.$ , while the 2H $'\rightarrow$ 2H$_b'$ (2H$'' \rightarrow$ 2H$_a^{''}$) distortion has an energy gain of only $3.1$ (6.6) meV/$f.u.$~.~Lower panel: 1T$'$-WTe$_2$ as a function of the PC ($f.u.$). Numbers in the crystal structure images represent atomic distances in~\AA. For the relation between fluence and PC see SM.}\label{fig1}
\end{figure*}

\begin{figure*}[t!]
\centering
\includegraphics[width=1\linewidth]{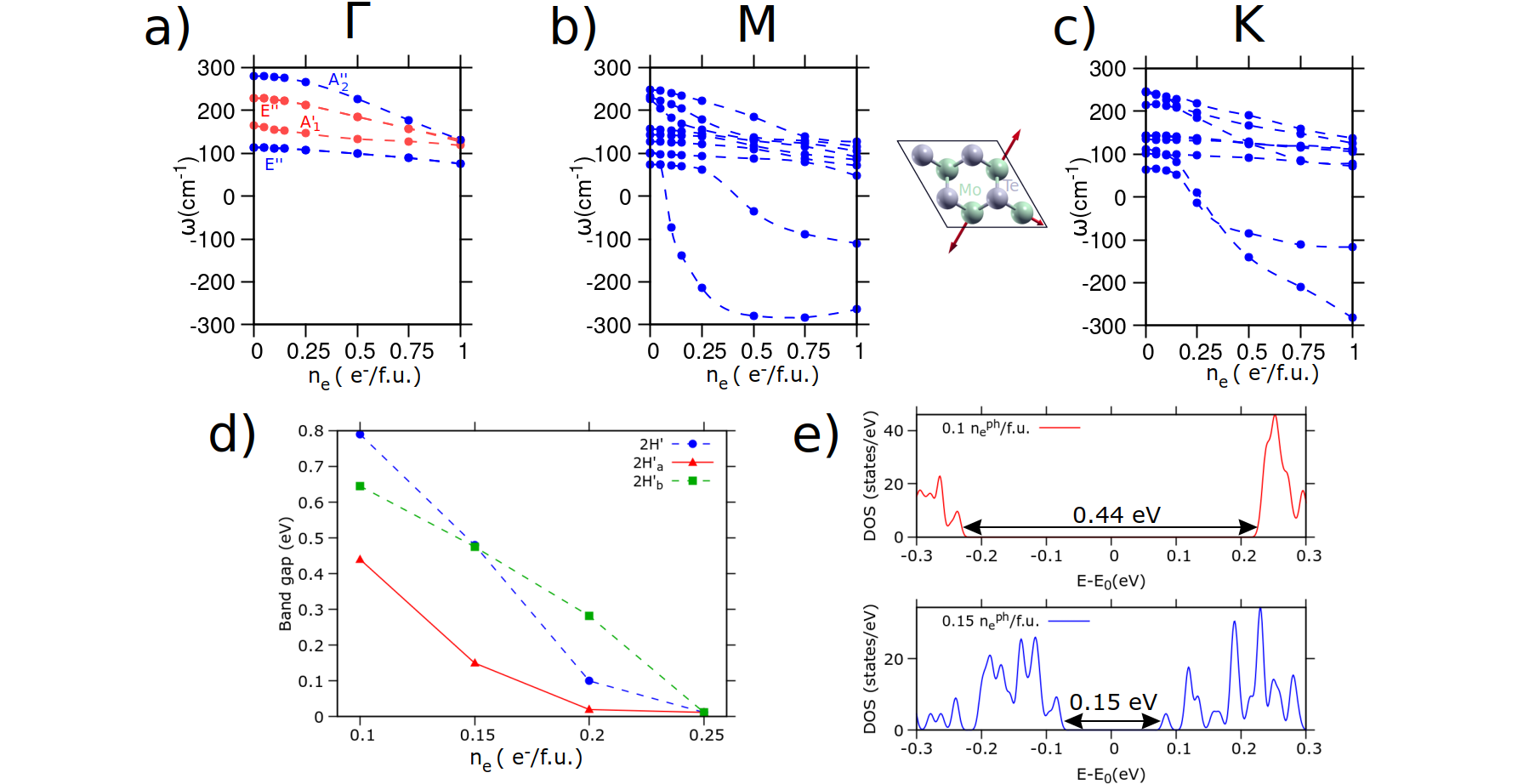}
\caption{ Panels a,b,c: Phonon frequencies as a function of PC for 2H-MoTe$_2$ at $\mathbf{\Gamma}$,$\mathbf{M}$,$\mathbf{K}$ points of the BZ, respectively. Raman active modes at $\mathbf{\Gamma}$ are depicted in red. The lowest eigenvector at the \textbf{M}-point is schematically depicted in panel b. Panel d: electronic bandgap as a function of PC. Panel e: density of states for the 2H$^\prime_a$ phase of MoTe$_2$ at $n_e=0.1$ and $0.15 ~e^-/f.u.$~.}\label{fig2}
\end{figure*}

The main results of our paper are summarized in Figs.~\ref{fig1},\ref{fig2}. First, we find that light induces a progressive formation of CDW order in 2H-MoTe$_2$ (and 2H-WTe$_2$), starting already at $n_e=0.1~e^-$ per formula unit ($f.u.$). 
The hidden CDW order, labeled 2H$^{\prime}$, arises from an imaginary phonon frequency in correspondence to the $\mathbf{M}$-point of the Brillouin zone (BZ) and involves clustering of three Mo atoms (the soft phonon pattern involves the displacement of Mo atoms only, see Fig.~\ref{fig2} (b)). The 2H$^{\prime}$ phase is in turn prone to structural instabilities, in particular two lower symmetry phases compete; we refer to them as 2H$^\prime_a$ and 2H$^\prime_b$. Most importantly, the 2H$^{\prime}$ phase, and the derived $a$ and $b$ distorted structures, display a progressive bandgap closing with increasing PC and transition to a metallic state at $n_e=0.25~e^-/f.u.$ (see Figs.~\ref{fig2} (d),(e)). At $n_e=1 ~e^-/f.u.$ (still reachable by current laser sources) a new $2\times 1$ CDW appears, composed of alternating anisotropically compressed and expanded stripes of hexagons. We label this phase 2H$^{''}$. The 2H$^{''}$ phase is in turn unstable and relaxes towards a lower symmetry structure, named 2H${''_a}$ phase.

Second, pumping on top of the 1T${^\prime}$-MoTe$_2$ and 1T${^\prime}$-WTe$_2$ phases at $n_e\approx 1~e^-/f.u.$ stabilizes a $2\times 2$ hidden CDW order composed of 2D diamond clusters of transition metal atoms (see Fig.~\ref{fig1} bottom right), that we label 1T${''}$. In the absence of optical pumping, this CDW structure has been found as metastable and competing with the 1T$^{\prime}$ phase in MoS$_2$\cite{PhysRevB.88.245428} and has been occasionally detected in some MoS$_2$ single layers\cite{PhysRevB.43.12053}. Thus, laser light can be used to reveal hidden charge density wave orders, hard to stabilize in standard thermochemical conditions. Finally, we note that energy transfer from the electrons to the lattice could reduce the effective temperature to be considered in the simulations, suggesting that the critical PC for the stabilization of the 1T$''$ phase is somewhat underestimated. 

 We now study the mechanism destabilizing the 2H structures in detail (for the analysis of the CDW formation in 1T$^{\prime}$ phases see SM). We consider  MoTe$_2$, but similar results for WTe$_2$ are shown in the SM. We perform supercell finite difference calculations of the phonon frequencies at specific high-symmetry points. 
The results are plotted in Fig.~\ref{fig2}~((a), (b), (c)). Raman active phonons at $\mathbf{\Gamma}$ are plotted in red. 
At low PCs, in a single-valley direct gap semiconductor, momentum and energy conservation enforce that  only zone center phonons can be excited.  Typically, if free internal coordinates are present, this results in displacive excitation of coherent Raman phonons. MoTe$_2$ is a two-valley semiconductor and phonon excitations at ${\bf \Gamma}$ and {\bf K} are both possible.
As it can be seen in Fig.\ref{fig2}~(a) the A$_1^{\prime}$ mode softens under weak photoexcitation. 
Recently it was suggested that phonon softenings at $\mathbf{\Gamma}$ drive the irreversible 2H-1T$^{\prime}$ transition in monolayer MoTe$_2$\cite{Peng2020}.~However, we find that even at large PCs $\mathbf{\Gamma}$ frequencies never become imaginary (Fig.~\ref{fig2}(a)), invalidating this claim. We find that a critical PC of $n_e \approx 1.75 ~e^-/f.u.$, corresponding to a high fluence of 31.41 mJ/cm$^{2}$, is  needed to destabilize $\mathbf{\Gamma}$ phonons, suggesting that other concomitant mechanisms such as Te vacancy creation are involved in the irreversible  2H$\to$1T$^{\prime}$transition~ \cite{doi:10.1021/acs.nanolett.9b00613}. At higher PCs instabilities can also occur at other phonon momenta. Already at $n_e=0.1~e^-/f.u.$, $\bf{M}$-point phonons are strongly imaginary. In order to understand the mechanism responsible for the formation of the 2H$'$ phases and to decouple the role of conduction and valence states in the CDW formation we also calculate the phonon frequency spectrum within a rigid doping approximation, simulating an electron charge excess or deficiency. The rigid doping calculations show that both conduction and valence states contribute to the $\bf{M}$-point phonon softening observed in cDFT (see Fig. S7 of SM).

As mentioned earlier, both 2H$^\prime$ and 2H$''$ phases present additional structural instabilities (see Figs. S8, S9 of SM). Concerning the 2H$^\prime$, we observe a competition between a 4$\times$4 (2H$^\prime_a$) and  2$\sqrt3\times2\sqrt3$ (2H$^\prime_b$) periodical distortion. The 2H$^\prime_a$ phase is more stable for $0.1 ~e^-/f.u. \leq n_e \leq 0.2 ~e^-/f.u.$~, while the 2H$^\prime_b$ is more stable at $n_e=0.25 ~e^-/f.u.$.
  At $n_e=1 ~e^-/f.u.$, the 2H$''$ instability (Fig. S5 of the SM, bottom right panel) leads to the 2H$''_a$ phase, having a 4$\times$2 periodicity. This further distortion produces an energy gain of 6.6 meV$/f.u.$ with respect to the 2H$''$ phase. 
We underline that the 2H$^{\prime}$ $\rightarrow$  2H$^\prime_b$ and 2H$''$ $\rightarrow$ 2H$''_a$ represent minor distortions relatively to the 2H $\rightarrow$ 2H$^{\prime}$ and 2H $\rightarrow$ 2H$''$ distortions respectively, as demonstrated by the relative energy gain: at $n_e=0.25 ~e^-/f.u.$ we calculate an energy gain of 107.2 meV$/f.u.$ for the 2H $\rightarrow$ 2H$^\prime$ distortion against an energy gain of 3.1 meV$/f.u.$ for the 2H$^{\prime}$ $\rightarrow$  2H$^{\prime}_b$ distortion; similarly, at $n_e=1 ~e^-/f.u.$  we calculate an energy gain of 583.1 meV$/f.u.$ for the 2H $\rightarrow$ 2H$''$ distortion against 6.6 meV$/f.u.$ for the 2H$''$ $\rightarrow$ 2H$_a''$ distortion. Whereas, this is not the case for the 2H$^\prime_a$ phase: at $n_e=0.15 ~e^-/f.u.$ we calculate comparable energy gains for the 2H $\rightarrow$ 2H$^\prime$ ($\approx$ 20 meV$/f.u.$) and the 2H$^{\prime}$ $\rightarrow$  2H$^{\prime}_a$ distortion ($\approx$ 29 meV$/f.u.$).

\begin{figure*}[t!]
\centering
\includegraphics[width=0.85\linewidth]{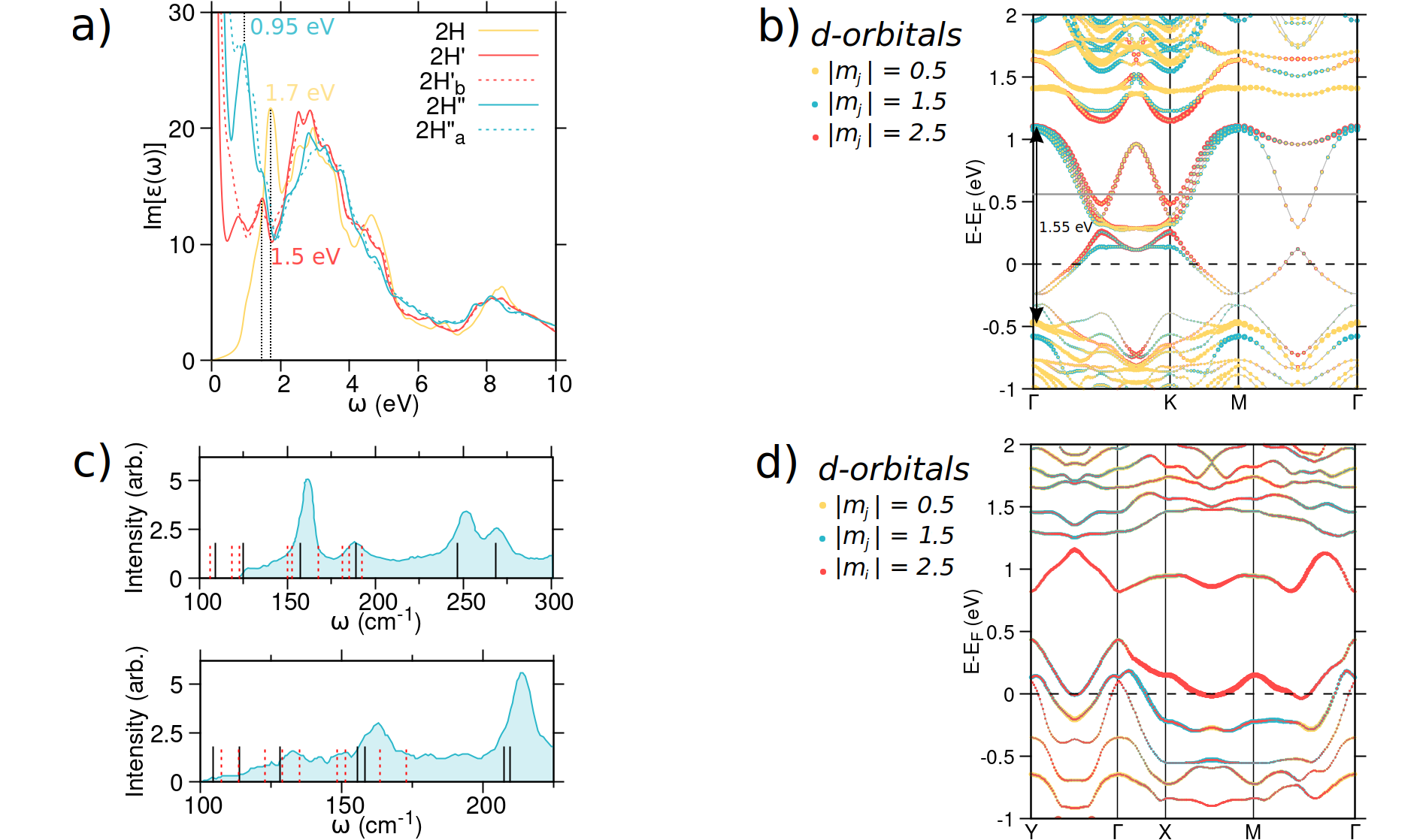}
\caption{Panel a: Imaginary part of $\epsilon(\omega)$ evaluated at $n_e=0, 0.25$ and $1 ~e^-/f.u.$ for 2H-MoTe$_2$ and its distorted phases. Panel b: Kohn-Sham eigenvalues along the 2H high symmetry BZ path for 2H$'$ MoTe$_2$ at $n_e = 0.25~e^-/f.u.$ , projected onto atomic $d$-orbitals. Black dashed line (grey line) represents the valence (conduction) $quasi$-Fermi level. 
Panel c: Calculated Raman frequencies (black lines) for 1T$'$-MoTe$_2$ (top) and WTe$_2$ (bottom), compared with the experimental Raman spectrum (cyan shaded region) for 1T$'$-MoTe$_2$\cite{doi:10.1021/acs.nanolett.6b01342} and 1T$'$-WTe$_2$\cite{Jiang2016}. Red dashed lines represent the Raman frequencies of the 1T$''$ phase at  $n_e=1~e^-/f.u.$. Panel d: Kohn-Sham eigenvalues along the 1T$'$ high symmetry BZ path for 1T$''$-WTe$_2$ at $n_e=1~e^-/f.u.$, projected onto atomic $d$-orbitals.}\label{fig3}
\end{figure*}

Having understood the mechanism leading to CDW formation, we identify possible spectroscopic signatures of the transient phases.
In the case of the 2H$^{\prime}$ and 2H$^{''}$ structures,
we calculate the imaginary part of the dielectric function $\epsilon(\omega)$ allowing both for valence to valence and conduction to conduction  optical transitions (absent in the ground state) on top of the valence to conduction ones.  In Fig.\ref{fig3} (a) , we consider both 2H$^\prime$, 2H$''$ phases (solid lines) and their respective distortions (dashed lines) at $n_e=0.25$ and $1 e^- /f.u.$ (see also Fig. S10 of SM).
For the ground state (i.e. no pumping), our results (see Figs. S11,S12 of SM) are in good agreement with previously published $ab$-$initio$ results\cite{2012PhyB..407.4627K}\footnote{In Ref.\cite{2012PhyB..407.4627K} the intensity of the monolayer dielectric function is incorrectly renormalized  due to the presence of the vacuum, an effect that in the paper is incorrectly attributed to the reduced number of bands.}.
The excited state configuration displays a metallic behaviour (Drude peak)  both in MoTe$_2$ and WTe$_2$ (see Fig.~\ref{fig3} (a) for MoTe$_2$ and Figs. S11-S14 of SM for WTe$_2$ as well as data on reflectance, energy loss and decomposition of Im$\{\epsilon(\omega)\}$ over optical transitions), which can be recognized from the low-frequency divergence of $\epsilon(\omega)$. This is related to the occurrence of  
an electron-hole plasma. The main changes in the dielectric function with increasing PC are observed below 2 eV. In particular, the peak observed in the ground state at 1.7 eV, attributed to the valence-conduction optical transition (Fig. S4 of SM, black arrow), is red-shifted to 1.5 eV at $n_e=0.25~e^-/f.u.$, as expected observing the corresponding electronic transition at the $\mathbf{\Gamma}$-point (see Fig.~\ref{fig3}~(b)). Furthermore, its intensity is notably reduced. At $n_e=1 ~e^-/f.u.$ this peak is absent, while a new peak at 0.95 eV not directly related to the valence-conduction contribution appears. Thus, monitoring the evolution of the peak detected at $1.7$ eV in the ground state as a function of the PC is equivalent to track the bandgap closing.
  The 2H$^\prime_a$, 2H$^\prime_b$ and 2H$''_a$ distortions slightly modify the optical response of the parent structures (see Fig. S10 of SM), causing a small red-shift of the 1.5 eV peak for the case of 2H$^\prime$ phase (down to 1.45 eV) and of the 0.95 eV peak (down to 0.89 eV) for the 2H$''$ phase and changing the Drude peak intensity.

For what concerns the 1T$^{\prime}$ and 1T$^{''}$ phases, a clear distinction between the two occurs in Raman spectra. The calculated Raman frequencies are shown in Fig.\ref{fig3}~(c) for 1T$^{\prime}$-MoTe$_2$ and 1T$^{\prime}$-WTe$_2$, and for photoexcited 1T$^{''}$-MoTe$_2$ and
1T$^{''}$-WTe$_2$. Those are compared to the experimental ground state Raman data from Refs.~\cite{Jiang2016,doi:10.1021/acs.nanolett.6b01342}.
 For 1T$^{\prime}$-MoTe$_2$, the agreement between theory and experiments is excellent. For the 1T$^{\prime}$-WTe$_2$ phase an overall red-shift ($\approx$ 2-3 \% of the value) of the phonon frequencies with respect to the experiment occurs, consistently with previous calculations\cite{Ma_2016} (see Tab.~S4 and the discussion in the SM). 
 We focus on the highest-energy most intense peaks (above $225$ cm$^{-1}$ for 1T$^{\prime}$-WTe$_2$ and above $200$ cm$^{-1}$ for 1T$^{\prime}$-MoTe$_2$), due to phonon modes of A$_{1g}$  and A$_{2g}$ symmetries. The two peaks are well separated in 
 1T$^{\prime}$-MoTe$_2$ and they are strongly overlapping in 1T$^{\prime}$-WTe$_2$, in good agreement with what found in experiments, demonstrating the reliability of our Raman calculations.
In the 1T$^{''}$ phase these two modes are completely absent. Thus, the structural transition is marked by the disappearance of these peaks in both compounds. The calculated Kohn-Sham eigenvalues for 1T$''$-WTe$_2$ are depicted in Fig.\ref{fig3}~(d): a completely different band structure with respect to 1T$'$-WTe$_2$ is observed (see also Figs. S17,S18 of SM).  

In conclusion, by using first principles calculations we have shown that light can be used to unveil hidden CDW order in the transient states of MoTe$_2$ and WTe$_2$ phases. Pumping on the 2H phases leads first to the stabilization of the 2H$^\prime_a$ CDW (4$\times$4 spatial periodicity), then to the 2H$^\prime_b$ CDW (2$\sqrt3\times2\sqrt3$ periodicity), and finally to the 2H$''_a$ phase (4$\times$2 periodicity). Most important, the  2H$'_a$ reconstruction shows an electronic gap that is progressively reduced as a function of fluence, showing a light tunable bandgap closing. Thus, inducing a photocarrier population in the 2H-MoTe$_2$ leads to the following successive transitions: 2H (insulating, no-CDW)$\to$  2H$'_a$ (gapped, 4$\times$4 CDW)$\to$ 2H$'_b$ (gapless, 2$\times$2 CDW)$\to$ 2H$''_a$ (gapless 4$\times$2). Given the similar results obtained for MoTe$_2$ and WTe$_2$, we speculate that similar 2H $\rightarrow$ 2H$'$ $\rightarrow$ 2H$''$ structural transformations occur in other 2D insulating dichalcogenides. We also show that pumping on the 1T$^{\prime}$ at $n_e=1 ~e^-/f.u.$ leads to the transient distorted 1T$''$ phase.  This new structure is characterized by diamond shaped Mo(W) clusters and a 2$\times$2 periodicity. For each hidden order we identify the spectroscopic fingerprints
paving the way to ultrafast measurements of transient states in monolayer dichalcogenides.

Finally, we have shown that the charge ordered phases occurring after laser
irradiation are different from those commonly observed in the ground state of metallic transition metal dichalcogenides, such as NbSe$_2$ or TiSe$_2$. Our work opens to the detection of new unexpected structural instabilities in this family of compounds.

\begin{acknowledgments}
We acknowledge support from the European Union's Horizon 2020 research and innovation programme Graphene Flagship under grant agreement No 881603. We acknowledge the CINECA award under the ISCRA initiative, for the availability of high performance computing resources and support. We acknowledge PRACE for awarding us access to Joliot-Curie at GENCI@CEA, France (project file number 2021240020).
\end{acknowledgments}

\bibliography{bibliography}
\bibliographystyle{apsrev4-2}
\end{document}